# Teaching Quantum Mechanical Commutation Relations via an Optical Experiment

A.Alper BİLLUR, Serkan AKKOYUN, Murat BURSAL

*Cumhuriyet Üniversitesi, Fen Fakültesi, Fizik Bölümü, Yüksek Enerji ve Plazma Fiziği Anabilim Dalı, 58140, Sivas, Turkey*

*Cumhuriyet Üniversitesi, Sağlık Hizmetleri Meslek Yüksek Okulu, Tıbbi Hizmetler ve Teknikleri Bölümü, 58140,Sivas, Turkey*

*Cumhuriyet Üniversitesi, Eğitim Fakültesi, İlköğretim Bölümü, Fen Bilgisi Eğitimi Anabilim Dalı, 58140, Sivas, Turkey*

**Abstract**

The quantum mechanical commutation relations, which are directly related to the Heisenberg uncertainty principle, have a crucial importance for understanding the quantum mechanics of students. During undergraduate level courses, the operator formalisms are generally given theoretically and it is documented that these abstract formalisms are usually misunderstood by the students. Based on the idea that quantum mechanical phenomena can be investigated via geometric optical tools, this study aims to introduce an experiment, where the quantum mechanical commutation relations are represented in a concrete way to provide students an easy and permanent learning. The experimental tools are chosen to be easily accessible and economic. The experiment introduced in this paper can be done with students or used as a demonstrative experiment in laboratory based or theory based courses requiring quantum physics content; particularly in physics, physics education and science education programs.

## Introduction

Despite the fact that Aristotle's description of physics in the history of science, one of the culminators of ancient Greek science, and also in the Islamic world as Muallim-i Evvel (First Teacher), was not supported by empirical evidence, he felt the influence of science as a prevailing physics vision for centuries. This widespread and effective understanding of physics can only be radically altered by Galileo, considered the first physicist in modern sense, and Newton, who is regarded as the summit of classical physics, in the 17th century called the Genius Age [1]. The Mathematical Principles of the Philosophy of Nature, which is one of Newton's most famous books of mathematical proof of the laws of attraction of planets using differential and integral calculus techniques in an elaborate manner, is a very important achievement of this new physics understanding [ 3]. It has been accepted that, with the contributions of important mathematicians of the 18th century such as Lagrange, Laplace, Paskal and Euler, the Newtonian mechanics became fully clear to every physical event in the world. This deterministic thought predicts that the universe operates like a machine, and that for any piece of this machine precise and clear predictions can be made at any time [1,4].

*Corresponding author: sakkoyun@cumhuriyet.edu.tr*

Events such as the creation of the periodic table in the 19th century and the calculation of the mathematical location and magnitude of this planet prior to the discovery of the planet Neptune by astronomers were accepted as final victories in terms of classical physics. Moreover, with these triumphs, it has begun to be claimed to have reached the final limits of physiology [1]. However, in the 20th century, the theory of quantum physics revealed by Planck's work on black body composition reveals different findings with many basic assumptions of classical physics.

The observable quantities (measurable) in classical physics are independent of the measurement of the observer. The classical mechanics, which states that the position and velocity of a particle can be measured at the same time and with definite accuracy, is deterministic in this sense. For example, using the Newtonian equations of motion, the position and momentum of a planet at a later time can be calculated with exact accuracy at the same time independently of the observer, if the solar system knows the initial position and velocity of the planet Mars. This way of thinking has been fundamentally changed in quantum space. Pictographically, it is not possible to determine the exact position and speed of electrons at the same exact time in an atomic system consisting of nuclei and electrons, similar to the solar system [5].

Experimental interpretations of quantum physics have been described in such cases as the ability of the classical physics to explain experimentally, the nature of light wave/particle duality, or the motion of atomic electrons. The Uncertainty Principle, an important milestone in the quantum physics, was introduced in 1927 contrary to the classic determinism of physics in this process [6,7]. According to this principle, two physical quantities related to each other, such as position and velocity, can not be measured at the same time with precise accuracy [8]. Although the uncertainty principle has been harshly criticized by many scientists including Einstein, in the 20th century successful developments in the field of quantum physics have been described as "the heart of the quantum physics" [9].

Many abstract concepts in quantum mechanics, including the uncertainty principle, lead to a perceived high level of difficulty in the quantum physics and modern physics courses given in this area. As a general conclusion, studies on this field [10] have shown that students tend to see quantum concepts difficult and incomprehensible in lessons that are taught using the traditional method of straight expression [11,12], although these concepts are not sufficiently effective [13] the learned concepts are not permanent [14,15,16].

Some researchers [15] point out that these learning deficits for quantum physics concepts are universal that students have difficulties in their lives, and that these difficulties are even found in students who have taken advanced quantum physics courses. For most students, the concept of quantum physics, including the Heisenberg uncertainty principle, is characterized by strange, mysterious, abstract mathematical formalism only and is very difficult to understand [11]. As a result, students are very low on these issues, under the influence of many misconceptions [15,17,18,19,20]. By avoiding these negative perceptions, it is recommended that these concepts be visualized and concretized as much as possible in order to provide more meaningful and lasting learning of quantum physics concepts [11,12,13,21,22,23].

**The Purpose and Importance of Research**

In the teaching of quantum mechanical concepts, optical experiments can be used to make use of concrete materials easier to apply [9]. However, when the related literature is examined it is seen that beyond the determination of the learning difficulties and causes related to the quantum physics, there are not enough examples of exploitation that can be used to overcome these problems [24] that many quantum mechanical phenomena can be made by optical experiments. In case studies in the literature on experimental demonstration of quantum mechanical commutation relations [25,26,27,28,29], these experiments have been found to be costly and difficult to access, therefore each student can not easily be done in the laboratory.

In this context, the aim of this work is to design an experiment in which geometric optics tools are used to teach quantum mechanical commutation relations, which are difficult to learn by university students and are directly related to the Heisenberg uncertainty principle, in a meaningful and permanent way. Economics and easy accessibility criteria are taken into consideration in the selection of experimental tools. It is thought that the experimental procedures described in this study can be used in demonstration experiment that can be done in all relevant laboratory applications, including physics, physics teacher and science teacher education programs in which quantum physics courses are given at all universities.

**Material and Method**

The experimental method designed within the scope of this study aims to teach concrete teaching of the quantum mechanical commutation relations in the courses of Introduction to Modern Physics and Quantum Physics given at the undergraduate level of universities. The experiment was designed by two experts trained in physics in nuclear physics and high energy physics. In order to get expert opinion, Prof. Abdullah Verçin was referred to who is the writer of a commonly used quantum mechanical textbook. In later chapters of the verse, experimental processes and mathematical formalism are described step by step, and these definitions and formalism are based on the textbook [8].

**Basic concepts**

a) Wavefunction: In quantum mechanics a state of a system is described by a complex wave function (Ψ(x)) that contains all the information about the systemd. The absolute square of the wave function gives the probability that the particle represented by this function is present in any desired position and at any given interval. Mathematically speaking, equation below gives the likelihood that the particle is located between points a and b in one dimension.

$$P = \int_a^b |\psi|^2 dx$$

For example, for an electron that can move at a certain speed, the wave function can be written in different forms in the position space and the momentum space. In fact, position and momentum observables are represented in two different spaces quantum mechanically, each of which is the Fourier transformation.

b) Operator: In quantum mechanics, the operators which operates on functions in specific spaces are mathematical expressions that transform the function to another. Operators are usually indicated by placing a (^) sign on them. For example, the derivation of the derivative function f is obtained by applying the derivative operator on f (x). The derivative operator can be represented mathematically as given by.

$$\hat{d}f(x) = f'(x)$$

The polarization operators (Â) exemplified in this study are the operators that transform the non-polarizing natural light to polarised light to a desired direction.

c) Uncertainty Relations: Since quantum mechanics is a theory of physics mostly related to microscopic particles, observables belonging to quantum mechanics are influenced by the measuring tools. Observables in quantum space are mathematically represented by Hermitian operators [30]. This phenomenon can be explained in the simplest way by the Heisenberg uncertainty principle [6]. For example, the uncertainties in position and momentum measurement at x coordinate can be represented by ($\Delta x$) and ($\Delta p_x$), respectively. Heisenberg uncertainty equation is given by.

$$\Delta x . \Delta p_x \geq \frac{\hbar}{2}$$

Here, the reduced Planck constant ($h/2\pi$) is the fundamental quantum of action with a value of $1.05457266 \times 10^{-34}$ joule.second. In the nature, no action can be observed that is smaller than this constant. According to above equation, the product of the uncertainties in the measurement of position and momentum can not be smaller than this fundamental constant. If this multiplication for any system is too large, the classical laws of physics are sufficient to explain the system. In contrast, if the order of multiplication is h that quantum physics laws are necessary to explain the system.

d) Commutation Relations: In a physics experiment, it may be desirable to take more than one measurement on the system. Different processors can be used for each measurement. Unlike classical mechanics, it is important to measure before and after for some operator groups in quantum space. The operation of these operators in the quantum mechanics in different orders on the wave function gives rise to different results as

$$\left[\hat{A},\hat{B}\right]\psi(x)=\left(\hat{A}.\hat{B}\right)\psi(x)-\left(\hat{B}.\hat{A}\right)\psi(x)\neq 0$$

Such processors may be referred "non-commuting operators". Whereas the commuting operators provide the condition

$$\left[\hat{A},\hat{B}\right]\psi(x)=\left(\hat{A}.\hat{B}\right)\psi(x)-\left(\hat{B}.\hat{A}\right)\psi(x)=0$$

The operating of these operators on the wave function in different orders gives same results.

The commutation relations can be exemplified in the macroscopic world as follows: A body located at any (a, b) coordinates on two axes perpendicular to each other which moves to (c, d) by subequent motions in two axes is independent of the order of operations of the transitions. So these different translations are called "commuting". Conversely, if the body is rotated around these two different axes, it depends on the order of rotation, and these rotations are called "non-commuting". There is a direct relationship between commutation relations and the above-mentioned Heisenberg uncertainty principle [31]. This relation can be mathematically expressed by

$$\left\langle \left(\Delta\hat{A}\right)^2 \right\rangle . \left\langle \left(\Delta\hat{B}\right)^2 \right\rangle \geq \frac{1}{4} \left| \left\langle \hat{A}.\hat{B} - \hat{B}.\hat{A} \right\rangle \right|^2$$

According to this expression, the lower limit of multiplication of uncertainty of two operators is related to whether the operators change order. Operators that change order with each other can be measured at the same time and with precise accuracy, but processors that can not change order with each other can not be measured with the same exact accuracy.

e) Polarization of Light: Light as an electromagnetic wave consists of a combination of electrical and magnetic fields, oscillating in perpendicular axes. The direction of propagation of the electromagnetic wave is perpendicular to both of the oscillation planes of these fields. Usually, the direction of polarization of the electromagnetic wave is defined as the direction of electric field vector (E). If the electric field component of the light emitted by a light source is oscillated in more than one direction, the light is called non-polarized (Figure 1.a). Besides, if there is a single direction of oscillation of the electric field component of the light, the light is called polarized light in a certain direction (Figure 1.c). The natural lights in daily life are non-polarized. In this study, the optical polarizers shown in Figure 1.b were used to polarize the light based on the selective absorption method.

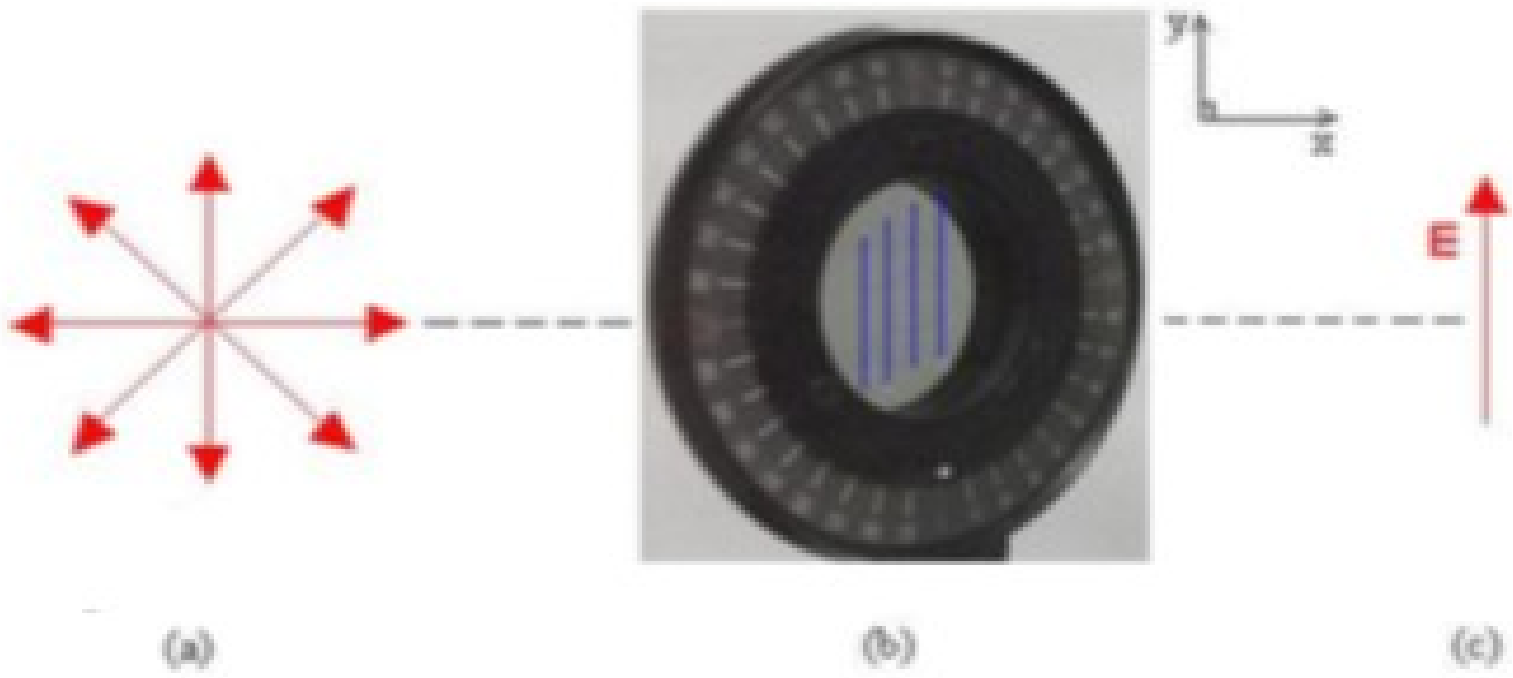

Figure 1. Polarization of non-polatized light (a) in y direction (c ) by using the polarizer (b) (Blue lines on the polarizer are represented)

The intensity of the beam that passed a polarizer represented by

$$I = I_0 \cos^2 \theta$$

where $I_0$ and I, respectively, the light intensities before and after passing polarizer. θ is the angle between the initial polarity of the light and the polarization axis. According to this law, called Malus

Law, the polarized light in a specific direction can not pass the polarizer which is located perpendicular to itself.

**Experimental Operations**

The apparatus is shown in Figure 2 which demonstrates the quantum mechanical commutation relations of physical operators to the students. In the apparatus, 630 to 680 nm wavelength light (electromagnetic wave) laser emitting diode, in 0.1 to 50,000 lux measuring interval with 0.1 lux resolution TT Technic VC1010 photometer and three light polarizer with adjustable polarizing angle. The polarizers are placed between the laser light source and the photometer as shown. All test steps were carried out in a dark environment where there was no light source other than the laser light source to keep the sensitivity of the experiment high. In case of the presence of external lights in the environment, background light substraction has to be performed.

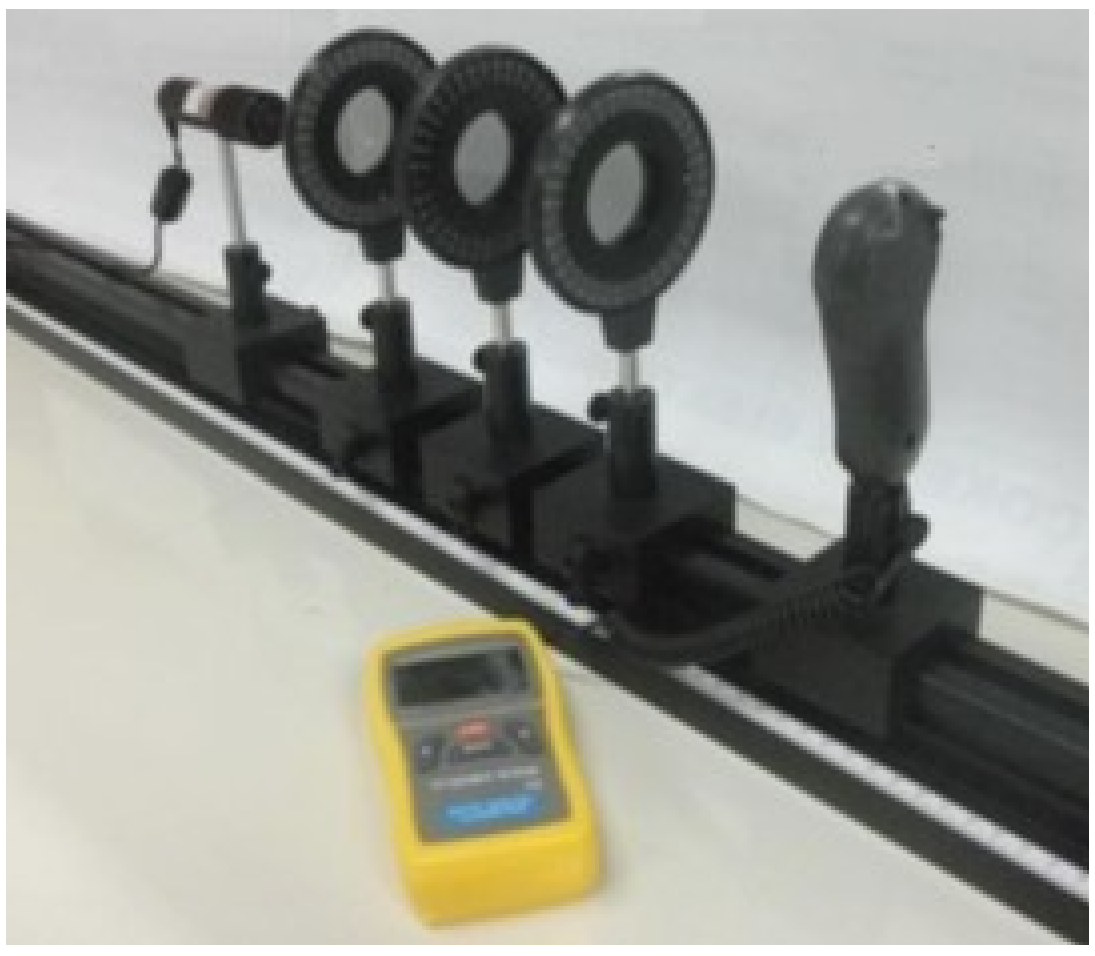

Figure 2. In the experimental setup in which representations of commutation relations are made, from left to right; (i) a light source diode laser, (ii) 3 polarizers, and (iii) a photometer

In the first step, between the light source and the light meter, two polarizers (polarizers $A_1$ and $A_3$) are placed. In this case, for each polarization angle set to be perpendicular to each other, the light from the source appears to have never reached the photometer. In the second step, a third polarizer (polarizer $A_2$) is placed between these two polarizers, such that the polarization angle is at a different angle than the other polarizers. In this case, it is seen that some light reaches the photometer.

In the system in which the light polarizers are represented as quantum physical processors, each polarizer is marked as a operator in order to show the commutation relations of these processors. Horizontally polarized ($0^o$) $A_1$ polarizer is represented by $\hat{A}_1$ opertor, vertical polarized (90°) $A_3$ polarizer is represented by $\hat{A}_3$ and diagonal polarized (45°) $A_2$ polarizer is represented by $\hat{A}_2$.

In Table 1, the luminosities are given in lux unit for different orders of polarizers between light source and photometer. The states are represented by wither "1" with some light reaching to photometer or "0" with no light reaching it. In Table 1, comparing luminosity values in row-1 and row-2 (or row-3), in row-1 there is light (23 lux) whereas in row-2 and 3 there is no light (0). The differences of row-2 (or 3) from row-1 is the commutation of $\hat{A}_2$ with $\hat{A}_3$ (or $\hat{A}_1$). This indicates that $\hat{A}_2$ can not commute with $\hat{A}_3$ (or $\hat{A}_1$). Mathematically

$$\left[\hat{A}_2,\hat{A}_3\right]\neq 0 \text{ veya } \left[\hat{A}_2,\hat{A}_1\right]\neq 0$$

The reason for this non-commutation relation is that the successive states of the operators placed at perpendicular angles to each other prevent the light from reaching the photometer.

On the other hand, with the experimental setup established above, they can also be shown to provide commutation relations. In Table 1, this can be observed by comparing the luminous intensities (23 and 106 lux) measured in rows-1 and 4. Differences between row-1 and 4 is $\hat{A}_1$ and $\hat{A}_3$ is replaced by each other. In both cases, the fact that some light reaches the light meter (states 1) indicates that these two processors can commute each other. By mathematical expression,

$$\left[\hat{A}_1,\hat{A}_3\right]=0$$

The presence of another operator with different angle (not 0 or 90°) placed between $\hat{A}_1$ and $\hat{A}_3$ which are perpendicular to each other, allows the certain amount of light to reach photometer in both cases.

In the study, quantum physically the polarization state of the light can be identified with the states of the spins 1/2 (fermion), although the commutation relations for the polarized and non-polarized states of a laser beam is examined. In this new case, the light polarizers are changed to the mechanisms that determine the spin orientation of the particles [28]. Such systems are called qubit (quantum bit) systems [31].

Table 1 In the case of three polarizers placed consecutively at different angles, luminous intensities from photometer

| No | Operators | | | Luminosity | State |
|---|---|---|---|---|---|
| Row-1 | $\hat{A}_1$ | $\hat{A}_2$ | $\hat{A}_3$ | 23 | 1 |
| Row-2 | $\hat{A}_1$ | $\hat{A}_3$ | $\hat{A}_2$ | 0 | 0 |
| Row-3 | $\hat{A}_2$ | $\hat{A}_1$ | $\hat{A}_3$ | 0 | 0 |
| Row-4 | $\hat{A}_3$ | $\hat{A}_2$ | $\hat{A}_1$ | 106 | 1 |

**Mathematical Representation**

The mathematical representation of the experimental results of commutation relations can be made using the Dirac notation. Accordingly, acting on non-polarized state the horizontal polarization operator $\hat{A}_1$ results horizontal polarized light state (|H>). Similarly, acting on non-polarized state the vertical polarization operator $\hat{A}_3$ results vertical polarized light state (|V>). Finally, acting on non-polarized state the diagonal polarization operator $\hat{A}_2$ results diagonal polarized light state (|H>+|V>). Additionally,

$$\hat{A}_1|Y\rangle = 1|Y\rangle \;\text{ ve }\; \hat{A}_1|D\rangle = 0|D\rangle$$

$$\hat{A}_3|Y\rangle = 0|Y\rangle \;\text{ ve }\; \hat{A}_3|D\rangle = 1|D\rangle$$

$$\hat{A}_2|Y\rangle = \frac{1}{\sqrt{2}}\left(|Y\rangle + |D\rangle\right) \;\text{ ve }\; \hat{A}_2|D\rangle = \frac{1}{\sqrt{2}}\left(|Y\rangle + |D\rangle\right)$$

Following these definitions, the mathematical proofs of the experimentally obtained results in Table 1 can be made as follows. As can be clearly seen from the operations carried out, in row-2 and row-3, no light would reach the photometer, while in the others were reached. This shows $\hat{A}_1$ and $\hat{A}_3$ commutes each other but they do not cummute with $\hat{A}_2$ .

**Row-1** $\hat{A}_1\hat{A}_2\hat{A}_3: \quad \rightarrow \hat{A}_1\hat{A}_2|D\rangle = \frac{1}{\sqrt{2}}\hat{A}_1\left(|Y\rangle + |D\rangle\right) = \frac{1}{\sqrt{2}}|Y\rangle$

**Row-2** $\hat{A}_1\hat{A}_3\hat{A}_2: \quad \rightarrow \frac{1}{\sqrt{2}}\hat{A}_1\hat{A}_3\left(|Y\rangle + |D\rangle\right) = \frac{1}{\sqrt{2}}\hat{A}_1|D\rangle = 0|D\rangle = 0$

**Row-3** $\hat{A}_2\hat{A}_1\hat{A}_3: \quad \rightarrow \hat{A}_2\hat{A}_1|D\rangle = \hat{A}_2\left(0|D\rangle\right) = 0$

**Row-4** $\hat{A}_3\hat{A}_2\hat{A}_1: \quad \rightarrow \hat{A}_3\hat{A}_2|Y\rangle = \frac{1}{\sqrt{2}}\hat{A}_3\left(|Y\rangle + |D\rangle\right) = \frac{1}{\sqrt{2}}|D\rangle$

**Conclusions and Recommendations**

In understanding quantum mechanics, the operator formalizations, commutation relations, and Heisenberg uncertainty principle have a basic prescription. But in the theoretical quantum mechanics lessons these concepts are often not understood sufficiently because of the limited use in the practice, using more traditional methods of expression. As a result, many courses in the literature supporting each other [11,12,13,14,15,16], these courses are considered to be very difficult in terms of students. Negative outlooks are being developed and very low achievements are exhibited. In this study, an experimental method which can be used to increase students' understanding of these basic concepts in quantum mechanics is proposed. Care has been taken to design the instrument used to carry out this experiment using materials that are readily accessible to instructional units.

**Acknowledgement**

Throughout our work, we thank Prof. Abdullah Verçin from Ankara University, Turkey that we have received his valuable opinions and suggestions.